\documentclass[aps,preprint,showpacs,preprintnumbers,amsmath,amssymb]{revtex4}
\usepackage{amsmath,mathrsfs,amsbsy,color,graphicx,bm,amsthm,amsfonts}
\usepackage{units}
\usepackage{bbm}
\usepackage{times}
\usepackage{dcolumn}
\usepackage{mathrsfs}
\usepackage{amsmath,amssymb,epsfig}
\begin{document}

\title{Does relativistic motion really freeze initially maximal entanglement?}
\author{Si-Han Li, Hui-Chen Yang, Rui-Yang Xu, Shu-Min Wu\footnote{Email: smwu@lnnu.edu.cn}}
\affiliation{ Department of Physics, Liaoning Normal University, Dalian 116029, China}


\begin{abstract}
We investigate the relativistic dynamics of quantum entanglement in a four-qubit cluster  ($CL_4$) state using a fully operational Unruh-DeWitt detector framework. Contrary to the widely held expectation that the Unruh effect inevitably degrades initially maximal entanglement, we demonstrate that the $1-3$ bipartite entanglement of the $CL_4$ state remains strictly maximal for all accelerations, including the infinite-acceleration limit.
This result uncovers a previously unexplored phenomenon, namely the ``complete freezing of initially maximal entanglement" under relativistic motion. To the best of our knowledge, this is the first identification and systematic characterization of such a phenomenon within a relativistic framework. These findings overturn the conventional view that acceleration universally diminishes maximal entanglement and establish the $CL_4$ state as a promising resource for quantum information processing in non-inertial or
curved-spacetime settings.
\end{abstract}

\vspace*{0.5cm}
\maketitle
\section{Introduction}
Quantum entanglement is a fundamental feature of quantum mechanics and constitutes a cornerstone of quantum information theory, enabling nonclassical correlations between spatially separated subsystems \cite{LSH1}. Among the various forms of multipartite entanglement, cluster states represent a particularly important and versatile class. Owing to their graph-based entanglement structure, cluster states naturally support scalable architectures and flexible connectivity, while allowing quantum information processing to be implemented solely through sequences of adaptive local measurements. These properties establish cluster states as indispensable resource states for measurement-based quantum computation \cite{LSH2,LSH3,LSH4,LSH5,LSH6,LSH7,LSH8,LSH9}.
Beyond their role in universal quantum computation, cluster states also constitute key resources for quantum communication networks, quantum simulation, and fault-tolerant quantum architectures. Their intrinsic resilience against noise, qubit loss, and operational imperfections facilitates their integration with quantum
error-correcting codes, thereby enhancing the stability and reliability of large-scale quantum information processing platforms \cite{LSH10,LSH11,LSH12,LSH13,LSH14}. From an experimental perspective, high-fidelity preparation and verification of cluster states on intermediate-scale quantum processors have been successfully demonstrated, underscoring their practical relevance and maturity as a resource for advanced quantum technologies  \cite{LSH15}.

Relativistic quantum information has emerged as a rapidly developing interdisciplinary field at the interface of quantum information theory, quantum field theory, and general relativity \cite{SDF1,SDF2,SDF3,SDF4,SDF5,SDF6,SDF7,SDF8,SDF9,SDF10,SDF11,SDF12,SDF13,SDF14,SDF15,SDF16,SDF17,SDF18,SDF19,SDF20,SDF21,SDF22,SDF23,SDF24,SDF25,SDF26,SDF27,SDF28,SDF29,SDF30,SDF31,SDF32,SDF33,SDF34,SDF35,SDF36,SDF37,SDF38,SDF39,SDF40,SDF41,SDF42,SDF43,SDF44,SDF45,SDF46,SDF47,SDF48,SDF49,SDF50,SDF51,SDF52,QYM1,QYM2,SDF53,SDF54,
AGL1,AGL2,AGL3,AGL4,AGL5,AGL6,AGL7,AGL8,AGL9,AGL10,AGL11,AGL12,AGL13,AGL14}. A central objective of this field is to elucidate how relativistic effects, such as the Unruh and Hawking effects, modify quantum resources initially encoded in entangled states, including entanglement, quantum steering, quantum discord, and quantum coherence \cite{SDF1,SDF2,SDF3,SDF4,SDF5,SDF6,SDF7,SDF8,SDF9,SDF10,SDF11,SDF12,SDF13,SDF14,SDF15,SDF16,SDF17,SDF18,SDF19,SDF20,SDF21,SDF22,SDF23,SDF24,SDF25,SDF26,SDF27,SDF28,SDF29,SDF30,SDF31,SDF32,SDF33,SDF34,SDF35,SDF36,SDF37,SDF38,SDF39,SDF40}. Within the single-mode approximation for free fields, a substantial body of work has indicated that relativistic effects generically lead to the degradation of these quantum resources. However, this approximation relies on a global field-mode decomposition and may not faithfully capture the physics of localized measurements or realistic detection processes, thereby limiting its operational relevance. To overcome these limitations, the Unruh-DeWitt detector model has been widely adopted as a more physically grounded framework \cite{SDF55,SDF56,SDF57,SDF58,SDF59,SDF60}. By modeling a localized two-level system interacting with a quantum field, this approach incorporates the operational aspects of detection while avoiding the nonlocal assumptions inherent in mode-based analyses. Studies based on the Unruh-DeWitt framework have revealed that the Unruh effect can lead to the complete destruction of bipartite quantum correlations and quantum coherence \cite{SDF61,SDF62,SDF63}, with analogous behavior observed in multipartite entangled states such as the GHZ and W states \cite{SDF64,SDF65}. Despite the prominent role of cluster states in quantum information processing, the relativistic behavior of the $CL_4$ state has remained largely unexplored. Given the irreversible degradation of quantum correlations induced by acceleration in simpler systems, it is therefore of considerable interest to examine whether the $CL_4$ state exhibits enhanced robustness under relativistic motion. In particular, we investigate whether this state can display an entanglement-freezing behavior, whereby entanglement remains invariant with increasing acceleration. Confirmation of such robustness would provide valuable insight into the preservation of quantum entanglement in relativistic settings and may inform the design of more resilient quantum information protocols in curved spacetime or non-inertial reference frames.

Motivated by the considerations outlined above, we investigate the behavior of quantum entanglement in the $CL_{4}$  state  within a relativistic tetrapartite system using the Unruh-DeWitt detector model. Initially, four observers-Alice, Bob, Charlie, and David share a $CL_{4}$ state in flat Minkowski spacetime. The detectors associated with Alice, Bob, and Charlie remain switched off and inertial, while David's detector undergoes uniform acceleration and interacts with a massless scalar field. We find that, in stark contrast to Bell, GHZ, and W states whose entanglement is completely degraded by the Unruh effect \cite{SDF61,SDF62,SDF63,SDF64,SDF65}, the entanglement of the $CL_{4}$ state remains strictly maximal for all accelerations, including the
infinite-acceleration limit. In other words, the entanglement is fully preserved throughout the relativistic evolution and remains entirely insensitive to the Unruh effect. This behavior constitutes a distinctive phenomenon, which we refer to as the complete freezing of initially maximal entanglement. To the best of our knowledge, this phenomenon is identified here for the first time within a relativistic framework.
The exceptional robustness of entanglement in the $CL_{4}$ state highlights its potential as a resilient quantum resource in relativistic settings and provides new insight into the interplay between quantum correlations and relativistic motion, with possible implications for quantum information protocols in high-acceleration regimes or curved spacetime backgrounds.

The structure of this paper is as follows. In Sec.~II, we introduce the quantum‐information framework for entangled Unruh-DeWitt detectors and study the evolution of the $CL_4$ state when one detector undergoes uniform acceleration. In Sec.~III, we examine the behavior of quantum entanglement in this relativistic setting. Finally, Sec.~IV summarizes our conclusions.

\section{Evolution of $CL_{4}$ state under Unruh-DeWitt detector model}

\begin{figure}[ht]
\centering
\includegraphics[width=0.9\textwidth]{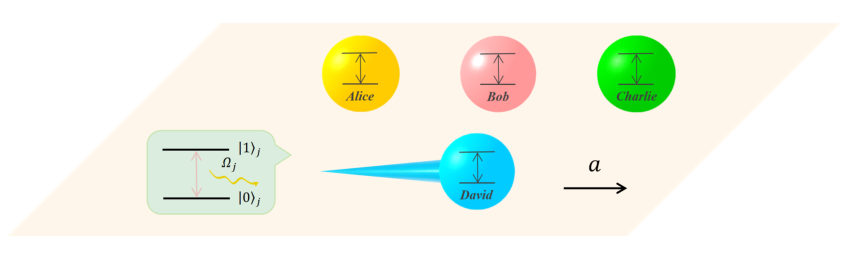}
\caption{Schematic illustration of the Unruh-DeWitt detector configuration. Four observers (Alice, Bob, Charlie, David) each carry an Unruh-DeWitt detector modeled as a two-level atom. The detectors of Alice, Bob and Charlie remain stationary, while David's detector undergoes uniform acceleration $a$ during a time interval $\Delta$.}
\label{Fig0}
\end{figure}

We consider four observers-Alice ($A$), Bob ($B$), Charlie ($C$) and David ($D$) each equipped with an Unruh-DeWitt detector modeled as a noninteracting two-level atom. The detectors of $A,B,C$ are inertial and remain switched off after preparation, whereas only David's detector is switched on and undergoes uniform proper acceleration $a$ during a finite proper-time interval $\Delta$ (see Fig.\ref{Fig0}). After this interval, David's detector continues its uniform acceleration while remaining coupled to the field. The worldline of David's detector is
\begin{eqnarray}\label{w66}
t(\tau)=a^{-1}\sinh a\tau, \quad x(\tau)=a^{-1}\cosh a\tau,
\end{eqnarray}
$y(\tau)=z(\tau)=0$, where $a$ is David's proper acceleration, and $\tau$ represents the proper time of the detector \cite{SDF61,SDF62}. Throughout this paper, we set $c=\hbar=k_{B}=1$.
The initial state of the detector-field system at time $t_{0}$ is taken to be a product of the four-qubit $CL_{4}$ state and the Minkowski vacuum,
\begin{eqnarray}\label{A7}
|\Psi^{ABCD\phi}_{t_{0}}\rangle=|\Psi_{ABCD}\rangle\otimes|0_{M}\rangle,
\end{eqnarray}
where the four-qubit $CL_{4}$ state is defined as
\begin{eqnarray}\label{qq7}
|\Psi_{ABCD}\rangle=\frac{1}{2}(|0_A0_B0_C0_D\rangle+|0_A0_B1_C1_D\rangle+|1_A1_B0_C0_D\rangle-|1_A1_B1_C1_D\rangle).
\end{eqnarray}
Here $|0_{M}\rangle$ denotes the Minkowski vacuum of the external scalar field, and $|0\rangle,|1\rangle$ represent the ground and excited states of each two-level detector, respectively.

The total Hamiltonian of the system is given by
\begin{eqnarray}\label{A9}
H_{4\phi}=H_{A}+H_{B}+H_{C}+H_{D}+H_{KG}+H_{\rm{int}}^{D\phi},
\end{eqnarray}
where $H_{T} = \Omega T^{\rm \dagger}T$ $(T=A,B,C,D)$ denotes the free Hamiltonian of the detectors with energy gap $\Omega$, $H_{KG}$ is the Hamiltonian of the massless scalar field, and $H_{\rm{int}}^{D\phi}$ describes the coupling between David's detector and the field. The ladder operators satisfy $T|1\rangle=|0\rangle$, $T^{\dagger}|0\rangle=|1\rangle$, and $T^{\dagger}|1\rangle=T|0\rangle=0$. We model the interaction by the usual Unruh-DeWitt coupling localized around the detector
\begin{eqnarray}\label{A8}
H_{\rm{int}}^{D\phi}(t)=\epsilon(t)\int_{\sum_{t}}d^{3} \mathbf{x}\sqrt{-g}\phi(x)[\chi(\mathbf{x})D+\bar{\chi}(\mathbf{x})D^{\rm \dagger}],
\end{eqnarray}
where $g\equiv \det(g_{ab})$, and $g_{ab}$ is the Minkowski spacetime metric. The function $\chi(\mathbf{x})=(\kappa\sqrt{2\pi})^{-3}\exp(-\mathbf{x}^{2}/(2\kappa^{2}))$ is a Gaussian coupling function, where the parameter $\kappa$ governs the effective size or interaction range of the detector. This profile vanishes outside a small region around the detector \cite{SDF61}, effectively modeling a pointlike detector interacting only with nearby field modes in the Minkowski vacuum.

In the weak-coupling regime, the final state $|\Psi^{D\phi}_{t=t_{0}+\Delta}\rangle$ of the atom-field system at time $t=t_{0}+\Delta$ can be calculated in first-order perturbation theory over the coupling constant $\epsilon$.  Under the evolution generated by the Hamiltonian in Eq.(\ref{A9}), the final state $|\Psi^{D\phi}_{t}\rangle$  at time
$t$ takes the form
\begin{eqnarray}\label{qq1}
|\Psi^{D\phi}_{t}\rangle=\{I-i[\phi(f)D+\phi(f)^{\rm \dagger}D^{\rm \dagger}]\}|\Psi^{D\phi}_{t_{0}}\rangle,
\end{eqnarray}
where the smeared field operator $\phi(f)$ is defined as
\begin{eqnarray}\label{qq2}
\begin{split}
\phi(f)\equiv&\int d^{4}x\sqrt{-g}\chi(x)f\\
=&i[a_{RI}(\overline{uE\overline{f}})-a^{\dagger}_{RI}(uEf)],
\end{split}
\end{eqnarray}
and describes the distribution of the external scalar field.
Here, $f\equiv\epsilon(t)e^{-i\Omega t}\chi(\mathbf{x})$ is a complex function with compact support in Minkowski spacetime, while $a_{RI}(\overline{u})$ and $a_{RI}^{\dagger}(u)$ represent the annihilation and creation operators of $u$ modes, respectively. The operator $u$ extracts the positive-frequency components of Klein-Gordon solutions in the Rindler metric, and  $E$ is the difference between the advanced and retarded Green's functions.

Substituting the initial state in Eq.(\ref{A7}) into Eq.(\ref{qq1}), we obtain the evolved state in terms of the Rindler operators $a^{\dagger}_{RI}$ and $a_{RI}$ as
\begin{eqnarray}\label{qqq2}
\begin{split}
|\Psi^{ABCD\phi}_{t}\rangle=&|\Psi^{ABCD\phi}_{t_{0}}\rangle+\frac{1}{2}[(|0010\rangle-|1110\rangle)\otimes(a^{\rm \dagger}_{RI}(\lambda)|0_{M}\rangle)\\
&+(|0001\rangle+|1101\rangle)\otimes(a_{RI}(\bar{\lambda})|0_{M}\rangle)],
\end{split}
\end{eqnarray}
where $\lambda=-uEf$, and the Rindler operators $a^{\dagger}_{RI}(\lambda)$ and $a_{RI}(\overline{\lambda})$ are defined in Rindler region $I$. The Minkowski vacuum is denoted by $|0_{M}\rangle$.
The Bogoliubov transformations between the Rindler operators and the operators annihilating the Minkowski vacuum state  are given by
\begin{eqnarray}\label{qq3}
a_{RI}(\bar{\lambda})=\frac{a_{M}(\overline{F_{1\Omega}})+e^{-\pi\Omega/a}a^{\rm \dagger}_{M}(F_{2\Omega})}{(1-e^{-2\pi\Omega/a})^{1/2}},
\end{eqnarray}
\begin{eqnarray}\label{qq5}
a^{\rm \dagger}_{RI}(\lambda)=\frac{a^{\rm \dagger}_{M}(F_{1\Omega})+e^{-\pi\Omega/a}a_{M}(\overline{F_{2\Omega}})}{(1-e^{-2\pi\Omega/a})^{1/2}},
\end{eqnarray}
where $F_{1\Omega}=\frac{\lambda+e^{-\pi\Omega/a}\lambda\circ w}{(1-e^{-2\pi\Omega/a})^{1/2}}$ and $F_{2\Omega}=\frac{\overline{\lambda\circ w}+e^{-\pi\Omega/a}\bar{\lambda}}{(1-e^{-2\pi\Omega/a})^{1/2}}$.
Here, the wedge reflection isometry $w(t,x,y,z)=(-t,-x,y,z)$ maps points in the Rindler region $I$ to region $II$, reflecting $\lambda$ to $\lambda\circ w$.

By applying the Bogoliubov transformations in Eqs.(\ref{qq3}) and (\ref{qq5}) along with the relations $a_{M}|0_{M}\rangle=0$ and $a_{M}^{\rm \dagger}|0_{M}\rangle=|1_{M}\rangle$, the evolved state in Eq.(\ref{qqq2}) can be recast in the form
\begin{eqnarray}\label{qqq5}
\begin{split}
|\Psi^{ABCD\phi}_{t}\rangle&=|\Psi^{ABCD\phi}_{t_{0}}\rangle+\frac{1}{2}\nu \left [ \frac{(|0010\rangle-|1110\rangle)
\otimes|1_{\tilde{F}_{1\Omega}}\rangle}{(1-e^{-2\pi\Omega/a})^{1/2}} \right. \\
& \quad \left. + e^{-\pi\Omega/a}\frac{(|0001\rangle+|1101\rangle)\otimes|1_{\tilde{F}_{2\Omega}}\rangle}{(1-e^{-2\pi\Omega/a})^{1/2}} \right],
\end{split}
\end{eqnarray}
where $\tilde{F}_{\rm{i}\Omega}=F_{\rm{i}\Omega}/\nu$.  To determine the state of the detectors after their interaction with the field, we trace out the external field degrees of freedom. This yields the reduced density matrix of the four-qubit system,
\begin{eqnarray}
\rho^{{ABCD}}_{t}=\|\Psi^{ABCD\phi}_{t}\|^{-2}{\rm{tr}}_{\phi}|\Psi^{ABCD\phi}_{t}\rangle\langle \Psi^{ABCD\phi}_{t}|,
\end{eqnarray}
where the normalization factor $\|\Psi^{ABCD\phi}_{t}\|^{2}$ is
$$\|\Psi^{ABCD\phi}_{t}\|^{2}=1+\frac{\nu^{2}(1+e^{-2\pi\Omega/a})}{2(1-e^{-2\pi\Omega/a})}.$$
Accordingly, the reduced density matrix of the detectors takes the form
\begin{eqnarray}\label{pp6}
\rho^{{ABCD}}_{t}=
 \left(\!\!\begin{array}{cccccccc}
\alpha&{\hskip 6pt} 0&{\hskip 6pt} 0&{\hskip 6pt} \alpha&{\hskip 6pt} \alpha&{\hskip 6pt} 0&{\hskip 6pt} 0&-\alpha\\
0&{\hskip 6pt} \beta&{\hskip 6pt} 0&{\hskip 6pt} 0&{\hskip 6pt} 0&{\hskip 6pt} \beta&{\hskip 6pt} 0&{\hskip 6pt} 0\\
0&{\hskip 6pt} 0&{\hskip 6pt} \gamma&{\hskip 6pt} 0&{\hskip 6pt} 0&{\hskip 6pt} 0&-\gamma&{\hskip 6pt} 0\\
\alpha&{\hskip 6pt} 0&{\hskip 6pt} 0&{\hskip 6pt} \alpha&{\hskip 6pt} \alpha&{\hskip 6pt} 0&{\hskip 6pt} 0&-\alpha\\
\alpha&{\hskip 6pt} 0&{\hskip 6pt} 0&{\hskip 6pt} \alpha&{\hskip 6pt} \alpha&{\hskip 6pt} 0&{\hskip 6pt} 0&-\alpha\\
0&{\hskip 6pt} \beta&{\hskip 6pt} 0&{\hskip 6pt} 0&{\hskip 6pt} 0&{\hskip 6pt} \beta&{\hskip 6pt} 0&0\\
0&{\hskip 6pt} 0&-\gamma&{\hskip 6pt} 0&{\hskip 6pt} 0&{\hskip 6pt} 0&{\hskip 6pt} \gamma&{\hskip 6pt} 0\\
-\alpha&{\hskip 6pt} 0&{\hskip 6pt} 0&-\alpha&-\alpha&{\hskip 6pt} 0&{\hskip 6pt} 0&{\hskip 6pt} \alpha\\
\end{array}\!\!\right) \textcolor{red}{.}
\end{eqnarray}
Although the full Hilbert space of the four-qubit system is $16$-dimensional ($2^4$), the specific structure of the initial state and the interaction Hamiltonian confines the system's evolution strictly to an $8$-dimensional subspace. Consequently, all matrix elements corresponding to basis states outside this subspace vanish identically. The matrix in Eq.(\ref{pp6}) is explicitly written in the ordered basis $|0000\rangle$, $|0001\rangle$, $|0010\rangle$, $|0011\rangle$, $|1100\rangle$, $|1101\rangle$, $|1110\rangle$ and $|1111\rangle$. Here, the matrix elements  $\alpha$, $\beta$, and $\gamma$ are given by
\begin{eqnarray}\label{l6}
\begin{aligned}
\alpha=\frac{1-q}{4(1-q)+2\nu^{2}(1+q)},\nonumber
\end{aligned}
\end{eqnarray}
\begin{eqnarray}\label{ll6}
\begin{aligned}
\beta=\frac{\nu^{2}q}{4(1-q)+2\nu^{2}(1+q)},\nonumber
\end{aligned}
\end{eqnarray}
\begin{eqnarray}\label{lll6}
\begin{aligned}
\gamma=\frac{\nu^{2}}{4(1-q)+2\nu^{2}(1+q)},\nonumber
\end{aligned}
\end{eqnarray}
where the acceleration parameter $q$ is defined as $q\equiv e^{-2\pi\Omega/a}$. The effective coupling strength is defined as $\nu^{2}\equiv\|\lambda\|^{2}=\frac{\epsilon^{2}\Omega\vartriangle}{2\pi}e^{-\Omega^{2}\kappa^{2}}$. For the validity of this definition, the condition $\Omega^{-1}\ll\vartriangle$ must hold. The parameter $\nu$ quantifies the effective interaction strength between the detector and the scalar field. Consequently, the condition $\nu^{2}\ll1$ corresponds to the weak-coupling limit, ensuring that the interaction energy is small relative to the intrinsic energy scales of the system. In this regime, the Dyson series expansion of the time-evolution operator allows for a valid truncation at the leading order, rendering higher-order contributions negligible. Thus, to ensure the validity of our perturbative approach, we strictly impose $\nu^{2}\ll1$ throughout the analysis. It is important to note that $q$ is a monotonous function of the acceleration $a$. Specifically, in the limit of zero acceleration, $q\rightarrow0$, and in the limit of infinite acceleration, $q\rightarrow1$ \cite{SDF62,SDF63}.

\section{ Quantum entanglement of the $CL_{4}$ state under the influence of Unruh thermal noise}
Negativity is a widely used entanglement measure that quantifies quantum correlations in both pure and mixed states. It is particularly effective for detecting entanglement because a bipartitioned density matrix
$\rho$  is entangled whenever its partial transpose exhibits at least one negative eigenvalue. For a four-partite (tetrapartite) system, the negativity associated with a given bipartition is defined as
\begin{eqnarray}\label{TH1}
N_{A(BCD)}=\|\rho^{{T_{A}}}_{ABCD}\|-1,
\end{eqnarray}
where $N_{A(BCD)}$ ( often referred to as the  $1-3$ tangle) quantifies the entanglement between subsystem $A$ and the remaining subsystems ($BCD$).
Here, $\|\rho^{{T_{A}}}_{ABCD}\|$  denotes the trace norm of the partially transposed density matrix. The notation $T_A$ indicates the partial transposition with respect to subsystem $A$ \cite{NE1}.
Alternatively, using the trace-norm definition  $\|O\|=tr\sqrt{O^{\dagger}O}$ for any Hermitian operator  $O$, the negativity may be rewritten in a form that is often more convenient for explicit computation
\begin{eqnarray}\label{TH2}
\|M\|-1=2\sum^{N}_{i=1}|\lambda^{(-)}_{M}|^{i},
\end{eqnarray}
where $|\lambda^{(-)}_{M}|^{i}$ denotes the absolute values of the negative eigenvalues of the partially transposed matrix $M$. This representation provides a direct and numerically efficient means of evaluating the entanglement associated with a given bipartition.

We now analyze the negativity, focusing on the $1-3$ tangle of the tetrapartite
$CL_{4}$ state described by the density matrix $\rho^{{ABCD}}_{t}$. Taking the partial transpose of $\rho^{{ABCD}}_{t}$ with respect to subsystem $A$, the operation maps elements out of the original invariant subspace. Therefore, we present $\rho^{{T_{A}}}_{ABCD}$ in the full $16$-dimensional computational basis ordered from $|0000\rangle$ to $|1111\rangle$:
\begin{eqnarray}\label{TH3}
\rho^{{T_{A}}}_{ABCD}=
\left(\!\!\begin{array}{ccccccccccccccccccc}
\alpha&{\hskip 6pt} 0&{\hskip 6pt} 0&{\hskip 6pt} \alpha&{\hskip 6pt} 0&{\hskip 6pt} 0&{\hskip 6pt} 0&{\hskip 6pt} 0&{\hskip 6pt} 0&{\hskip 6pt} 0&{\hskip 6pt} 0&{\hskip 6pt} 0&{\hskip 6pt} 0&{\hskip 6pt} 0&{\hskip 6pt} 0&{\hskip 6pt} 0\\
0&{\hskip 6pt} \beta&{\hskip 6pt} 0&{\hskip 6pt} 0&{\hskip 6pt} 0&{\hskip 6pt} 0&{\hskip 6pt} 0&{\hskip 6pt} 0&{\hskip 6pt} 0&{\hskip 6pt} 0&{\hskip 6pt} 0&{\hskip 6pt} 0&{\hskip 6pt} 0&{\hskip 6pt} 0&{\hskip 6pt} 0&{\hskip 6pt} 0\\
0&{\hskip 6pt} 0&{\hskip 6pt} \gamma&{\hskip 6pt} 0&{\hskip 6pt} 0&{\hskip 6pt} 0&{\hskip 6pt} 0&{\hskip 6pt} 0&{\hskip 6pt} 0&{\hskip 6pt} 0&{\hskip 6pt} 0&{\hskip 6pt} 0&{\hskip 6pt} 0&{\hskip 6pt} 0&{\hskip 6pt} 0&{\hskip 6pt} 0\\
\alpha&{\hskip 6pt} 0&{\hskip 6pt} 0&{\hskip 6pt} \alpha&{\hskip 6pt} 0&{\hskip 6pt} 0&{\hskip 6pt} 0&{\hskip 6pt} 0&{\hskip 6pt} 0&{\hskip 6pt} 0&{\hskip 6pt} 0&{\hskip 6pt} 0&{\hskip 6pt} 0&{\hskip 6pt} 0&{\hskip 6pt} 0&{\hskip 6pt} 0\\
0&{\hskip 6pt} 0&{\hskip 6pt} 0&{\hskip 6pt} 0&{\hskip 6pt} 0&{\hskip 6pt} 0&{\hskip 6pt} 0&{\hskip 6pt} 0&{\hskip 6pt} \alpha&{\hskip 6pt} 0&{\hskip 6pt} 0&{\hskip 6pt} \alpha&{\hskip 6pt} 0&{\hskip 6pt} 0&{\hskip 6pt} 0&{\hskip 6pt} 0\\
0&{\hskip 6pt} 0&{\hskip 6pt} 0&{\hskip 6pt} 0&{\hskip 6pt} 0&{\hskip 6pt} 0&{\hskip 6pt} 0&{\hskip 6pt} 0&{\hskip 6pt} 0&{\hskip 6pt} \beta&{\hskip 6pt} 0&{\hskip 6pt} 0&{\hskip 6pt} 0&{\hskip 6pt} 0&{\hskip 6pt} 0&{\hskip 6pt} 0\\
0&{\hskip 6pt} 0&{\hskip 6pt} 0&{\hskip 6pt} 0&{\hskip 6pt} 0&{\hskip 6pt} 0&{\hskip 6pt} 0&{\hskip 6pt} 0&{\hskip 6pt} 0&{\hskip 6pt} 0& -\gamma&{\hskip 6pt} 0&{\hskip 6pt} 0&{\hskip 6pt} 0&{\hskip 6pt} 0&{\hskip 6pt} 0\\
0&{\hskip 6pt} 0&{\hskip 6pt} 0&{\hskip 6pt} 0&{\hskip 6pt} 0&{\hskip 6pt} 0&{\hskip 6pt} 0&{\hskip 6pt} 0& -\alpha&{\hskip 6pt} 0&{\hskip 6pt} 0& -\alpha&{\hskip 6pt} 0&{\hskip 6pt} 0&{\hskip 6pt} 0&{\hskip 6pt} 0\\
0&{\hskip 6pt} 0&{\hskip 6pt} 0&{\hskip 6pt} 0&{\hskip 6pt} \alpha&{\hskip 6pt} 0&{\hskip 6pt} 0& -\alpha&{\hskip 6pt} 0&{\hskip 6pt} 0&{\hskip 6pt} 0&{\hskip 6pt} 0&{\hskip 6pt} 0&{\hskip 6pt} 0&{\hskip 6pt} 0&{\hskip 6pt} 0\\
0&{\hskip 6pt} 0&{\hskip 6pt} 0&{\hskip 6pt} 0&{\hskip 6pt} 0&{\hskip 6pt} \beta&{\hskip 6pt} 0&{\hskip 6pt} 0&{\hskip 6pt} 0&{\hskip 6pt} 0&{\hskip 6pt} 0&{\hskip 6pt} 0&{\hskip 6pt} 0&{\hskip 6pt} 0&{\hskip 6pt} 0&{\hskip 6pt} 0\\
0&{\hskip 6pt} 0&{\hskip 6pt} 0&{\hskip 6pt} 0&{\hskip 6pt} 0&{\hskip 6pt} 0& -\gamma&{\hskip 6pt} 0&{\hskip 6pt} 0&{\hskip 6pt} 0&{\hskip 6pt} 0&{\hskip 6pt} 0&{\hskip 6pt} 0&{\hskip 6pt} 0&{\hskip 6pt} 0&{\hskip 6pt} 0\\
0&{\hskip 6pt} 0&{\hskip 6pt} 0&{\hskip 6pt} 0&{\hskip 6pt} \alpha&{\hskip 6pt} 0&{\hskip 6pt} 0& -\alpha&{\hskip 6pt} 0&{\hskip 6pt} 0&{\hskip 6pt} 0&{\hskip 6pt} 0&{\hskip 6pt} 0&{\hskip 6pt} 0&{\hskip 6pt} 0&{\hskip 6pt} 0\\
0&{\hskip 6pt} 0&{\hskip 6pt} 0&{\hskip 6pt} 0&{\hskip 6pt} 0&{\hskip 6pt} 0&{\hskip 6pt} 0&{\hskip 6pt} 0&{\hskip 6pt} 0&{\hskip 6pt} 0&{\hskip 6pt} 0&{\hskip 6pt} 0&{\hskip 6pt} \alpha&{\hskip 6pt} 0&{\hskip 6pt} 0& -\alpha\\
0&{\hskip 6pt} 0&{\hskip 6pt} 0&{\hskip 6pt} 0&{\hskip 6pt} 0&{\hskip 6pt} 0&{\hskip 6pt} 0&{\hskip 6pt} 0&{\hskip 6pt} 0&{\hskip 6pt} 0&{\hskip 6pt} 0&{\hskip 6pt} 0&{\hskip 6pt} 0&{\hskip 6pt} \beta&{\hskip 6pt} 0&{\hskip 6pt} 0\\
0&{\hskip 6pt} 0&{\hskip 6pt} 0&{\hskip 6pt} 0&{\hskip 6pt} 0&{\hskip 6pt} 0&{\hskip 6pt} 0&{\hskip 6pt} 0&{\hskip 6pt} 0&{\hskip 6pt} 0&{\hskip 6pt} 0&{\hskip 6pt} 0&{\hskip 6pt} 0&{\hskip 6pt} 0&{\hskip 6pt} \gamma&{\hskip 6pt} 0\\
0&{\hskip 6pt} 0&{\hskip 6pt} 0&{\hskip 6pt} 0&{\hskip 6pt} 0&{\hskip 6pt} 0&{\hskip 6pt} 0&{\hskip 6pt} 0&{\hskip 6pt} 0&{\hskip 6pt} 0&{\hskip 6pt} 0&{\hskip 6pt} 0& -\alpha&{\hskip 6pt} 0&{\hskip 6pt} 0&{\hskip 6pt} \alpha\\
\end{array}\!\!\right).
\end{eqnarray}
Using Eq.(\ref{TH2}), the negativity associated with the bipartition  $N_{A(BCD)}$ is found to be
\begin{eqnarray}\label{TH55}
N_{A(BCD)}=1,
\end{eqnarray}
indicating maximal entanglement between subsystem $A$ and the remaining parties. Owing to the symmetry between Alice and Bob in the definition of the $CL_{4}$ state [Eq.(\ref{pp6})], it immediately follows that $N_{A(BCD)}=N_{B(ACD)}$. Consequently, the entanglement between Alice (or Bob) and the rest of the system remains strictly maximal throughout the entire range of accelerations. This behavior constitutes an instance of the ``complete freezing of initially maximal entanglement" in a relativistic setting. Remarkably, such robustness is absent in conventional multipartite states, including GHZ and W states, whose entanglement is known to undergo significant degradation with increasing acceleration \cite{SDF64,SDF65}. Our result therefore challenges the widely held expectation that the Unruh effect inevitably diminishes initially maximal entanglement.

We next evaluate the entanglement across the remaining $1-3$ bipartitions by performing the partial transposition with respect to modes $C$ and $D$. The resulting partially transposed density matrices, $\rho^{{T_{C}}}_{ABCD}$ and $\rho^{{T_{D}}}_{ABCD}$, can be expressed in the same ordered basis $|0000\rangle$, $|0001\rangle$, $|0010\rangle$, $|0011\rangle$, $|1100\rangle$, $|1101\rangle$, $|1110\rangle$ and $|1111\rangle$ as Eq.(\ref{pp6}):
\begin{eqnarray}\label{TH333}
\rho^{{T_{C}}}_{ABCD}=
 \left(\!\!\begin{array}{cccccccc}
\alpha&{\hskip 6pt} 0&{\hskip 6pt} 0&{\hskip 6pt} 0&{\hskip 6pt} \alpha&{\hskip 6pt} 0&{\hskip 6pt} 0&{\hskip 6pt} 0\\
0&{\hskip 6pt} \beta&{\hskip 6pt} \alpha&{\hskip 6pt} 0&{\hskip 6pt} 0&{\hskip 6pt} \beta&{\hskip 6pt} \alpha&{\hskip 6pt} 0\\
0&{\hskip 6pt} \alpha&{\hskip 6pt} \gamma&{\hskip 6pt} 0&{\hskip 6pt} 0&-\alpha&-\gamma&{\hskip 6pt} 0\\
0&{\hskip 6pt} 0&{\hskip 6pt} 0&{\hskip 6pt} \alpha&{\hskip 6pt} 0&{\hskip 6pt} 0&{\hskip 6pt} 0&-\alpha\\
\alpha&{\hskip 6pt} 0&{\hskip 6pt} 0&{\hskip 6pt} 0&{\hskip 6pt} \alpha&{\hskip 6pt} 0&{\hskip 6pt} 0&{\hskip 6pt} 0\\
0&{\hskip 6pt} \beta&-\alpha&{\hskip 6pt} 0&{\hskip 6pt} 0&{\hskip 6pt} \beta&-\alpha&{\hskip 6pt} 0\\
0&{\hskip 6pt} \alpha&-\gamma&{\hskip 6pt} 0&{\hskip 6pt} 0&-\alpha&{\hskip 6pt} \gamma&{\hskip 6pt} 0\\
0&{\hskip 6pt} 0&{\hskip 6pt} 0&-\alpha&{\hskip 6pt} 0&{\hskip 6pt} 0&{\hskip 6pt} 0&{\hskip 6pt} \alpha\\
\end{array}\!\!\right),
\end{eqnarray}
\begin{eqnarray}\label{TH4}
\rho^{{T_{D}}}_{ABCD}=
 \left(\!\!\begin{array}{cccccccc}
\alpha&{\hskip 6pt} 0&{\hskip 6pt} 0&{\hskip 6pt} 0&{\hskip 6pt} \alpha&{\hskip 6pt} 0&{\hskip 6pt} 0&{\hskip 6pt} 0\\
0&{\hskip 6pt} \beta&{\hskip 6pt} \alpha&{\hskip 6pt} 0&{\hskip 6pt} 0&{\hskip 6pt} \beta&-\alpha&{\hskip 6pt} 0\\
0&{\hskip 6pt} \alpha&{\hskip 6pt} \gamma&{\hskip 6pt} 0&{\hskip 6pt} 0&{\hskip 6pt} \alpha&-\gamma&{\hskip 6pt} 0\\
0&{\hskip 6pt} 0&{\hskip 6pt} 0&{\hskip 6pt} \alpha&{\hskip 6pt} 0&{\hskip 6pt} 0&{\hskip 6pt} 0&-\alpha\\
\alpha&{\hskip 6pt} 0&{\hskip 6pt} 0&{\hskip 6pt} 0&{\hskip 6pt} \alpha&{\hskip 6pt} 0&{\hskip 6pt} 0&{\hskip 6pt} 0\\
0&{\hskip 6pt} \beta&{\hskip 6pt} \alpha&{\hskip 6pt} 0&{\hskip 6pt} 0&{\hskip 6pt} \beta&-\alpha&{\hskip 6pt} 0\\
0&-\alpha&-\gamma&{\hskip 6pt} 0&{\hskip 6pt} 0&-\alpha&{\hskip 6pt} \gamma&{\hskip 6pt} 0\\
0&{\hskip 6pt} 0&{\hskip 6pt} 0&-\alpha&{\hskip 6pt} 0&{\hskip 6pt} 0&{\hskip 6pt} 0&{\hskip 6pt} \alpha\\
\end{array}\!\!\right).
\end{eqnarray}
By employing Eq.(\ref{TH2}), the corresponding negativities  (the  $1-3$ tangles) for  $N_{C(ABD)}$ and $N_{D(ABC)}$  are obtained as
\begin{eqnarray}\label{TH555}
N_{C(ABD)}=\max\left\{0,\frac{2-2q}{2-2q+\nu^{2}+q\nu^{2}}\right\},
\end{eqnarray}
\begin{eqnarray}\label{TH5}
N_{D(ABC)}=\max\left\{0,\frac{-\nu^{2}-q\nu^{2}+\sqrt{4-8q+4q^{2}+\nu^{4}-2q\nu^{4}+q^{2}\nu^{4}}}{2-2q+\nu^{2}+q\nu^{2}}\right\}.
\end{eqnarray}

\begin{figure}
\begin{minipage}[t]{0.5\linewidth}
\centering
\includegraphics[width=3.0in,height=6.24cm]{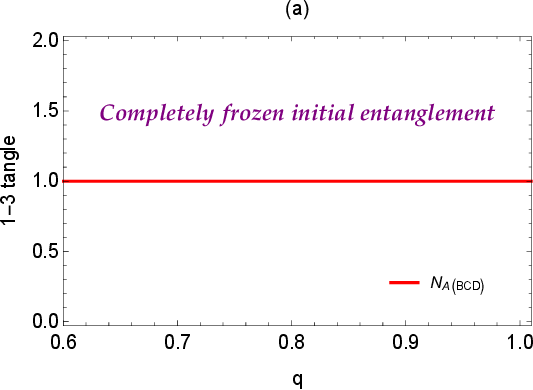}
\label{fig1a}
\end{minipage}%
\begin{minipage}[t]{0.5\linewidth}
\centering
\includegraphics[width=3.0in,height=6.24cm]{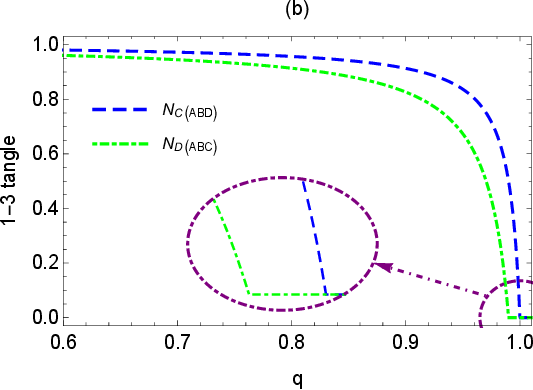}
\label{fig1b}
\end{minipage}%
\caption{The $1-3$ tangle of $CL_{4}$ state as a function of the acceleration parameter $q$. The effective coupling parameter is fixed as $\nu^2=0.01$. Note that the strictly constant value of $1$ for $N_{A(BCD)}$ represents an exact result within the framework of first-order perturbation theory, valid under the weak-coupling condition $\nu^2 \ll 1$.}
\label{Fig1}
\end{figure}

Fig.\ref{Fig1} illustrates the behavior of the $1-3$ tangle (negativity) of the tetrapartite $CL_{4}$ state as a function of the acceleration parameter $q$, with the effective coupling fixed at $\nu^2 = 0.01$. Remarkably, the negativity $N_{A(BCD)}$ remains identically equal to its maximal value over the entire acceleration range, showing no sensitivity to Unruh-induced thermal noise. This provides clear evidence of the phenomenon that we refer to as the ``complete freezing of initially maximal entanglement".
It is important to note that the result $N_{A(BCD)}=1$ is derived within the framework of first-order perturbation theory, which requires the weak-coupling condition $\nu^2 \ll 1$. While the ``complete freezing" phenomenon appears as a straight line in Fig.\ref{Fig1}, strictly speaking, this exact invariance holds within the validity of the perturbative approximation. In the limit of infinite acceleration ($q \to 1$), although higher-order contributions (scaling as $\mathcal{O}(\nu^4)$) might introduce minute corrections to the entanglement value, these corrections are negligible in the weak-coupling regime. Therefore, the essential physical feature---the exceptional robustness of the $CL_4$ state against the Unruh effect---remains valid.

It should be noted that in the specific configuration studied, all two-particle entanglement measures (such as \( N_{AB} \), \( N_{AC} \), \( N_{AD} \), etc.) vanish under relativistic evolution. Therefore, the global entanglement between subsystem \( A \) and the remaining three subsystems (\( BCD \)), namely \( N_{A(BCD)} \), satisfies:
\[
\pi_A = N_{A(BCD)}^2 - N_{AB}^2 - N_{AC}^2 - N_{AD}^2 = N_{A(BCD)}^2,
\]
which reflects the highly global and indivisible nature of the four-partite entanglement structure. In the \( CL_4 \) state, entanglement is not simply stored between any two particles, but is distributed over the connectivity graph of the entire cluster. Specifically, \( A \) is connected to \( B \), \( C \), and \( D \) via entanglement, yet the reduced density matrix of any two-qubit subsystem is separable, hence the bipartite entanglement is zero. However, the entanglement between \( A \) and the entire \( BCD \) subsystem remains maximal, demonstrating the typical ``global entanglement'' property of cluster states.
In the physical scenario considered, only David's (\( D \)) detector undergoes acceleration and couples to the external field, while \( A \), \( B \), and \( C \) remain inertial with their detectors switched off. The Unruh effect, acting as local thermal noise, directly affects the accelerated detector \( D \) and the field modes it couples to. Nevertheless, due to the highly nonlocal and symmetric entanglement structure of the \( CL_4 \) state, the entanglement between \( A \) and the entire \( BCD \) subsystem is not reliant on any specific bipartite channel (e.g., \( A \)-\( B \) or \( A \)-\( D \)), but is encoded in particular coherent superpositions of the four-partite system. These terms are protected during evolution by the overall symmetry of the system and the structure of the Hamiltonian, so that even though \( D \) experiences acceleration and introduces decoherence, the global entanglement between \( A \) and \( BCD \) remains completely preserved.

Physically, this result implies that the quantum entanglement between a single inertial qubit and the remaining tripartite subsystem is perfectly preserved under relativistic motion, exhibiting no degradation even in the infinite-acceleration limit.
Such behavior is highly nontrivial and stands in sharp contrast to the typical scenario encountered in relativistic quantum information. For instance, the entanglement of GHZ states is known to be extremely fragile, often suffering from ``sudden death" under relativistic noise due to its reliance on global coherence across all parties \cite{SDF64}. Similarly, while W states are generally more robust against particle loss, their entanglement remains susceptible to acceleration-induced degradation, as the noise on one partition directly affects the shared excitation \cite{SDF65}. In contrast to these canonical states, the $CL_4$ state possesses a linear graph structure where correlations are localized between neighbors. This distinctive topological feature ensures that the dominant correlations contributing to $N_{A(BCD)}$ remain structurally isolated from the thermal noise acting on David. The complete freezing observed here therefore underscores a specific advantage of cluster-state architectures over GHZ and W states in relativistic settings.

From an operational perspective, the observed ``complete freezing" of entanglement is of significant importance as it provides a robust quantum resource that is immune to relativistic degradation. In relativistic quantum information processing, entanglement is the fundamental resource for protocols such as quantum teleportation. Typically, the Unruh effect acts as environmental noise that depletes this resource, thereby reducing the fidelity of information transfer. However, the invariance of the maximal entanglement in the $CL_4$ state ensures that the quantum channel capacity is preserved. Consequently, this guarantees that the teleportation fidelity between the inertial observer and the remaining subsystems remains optimal and strictly surpasses the classical limit, regardless of the fourth party's acceleration. This exceptional stability establishes the $CL_4$ state as a superior candidate for implementing reliable quantum communication protocols in non-inertial frames, overcoming the decoherence typically induced by the Unruh effect.

More broadly, it reveals that certain structured multipartite states, exemplified by the $CL_{4}$ cluster state, possess an intrinsic robustness against relativistic noise. From an applied viewpoint, this robustness has important implications for relativistic quantum technologies. It suggests that cluster-type multipartite entangled states can function as reliable quantum resources in non-inertial frames or strong-gravity environments, enabling quantum communication and information-processing tasks under extreme spacetime conditions. Potential applications include relativistic quantum communication protocols, satellite-based quantum networks, and quantum information processing in high-acceleration or high-gravity regimes. Overall, these findings not only deepen our understanding of quantum correlations in relativistic settings but also open new avenues for preserving and exploiting quantum resources beyond the inertial domain.

On the other hand, the negativities $N_{C(ABD)}$ and $N_{D(ABC)}$, which involve the inertial observer Charlie and the accelerated observer David, exhibit qualitatively distinct behaviors. As the acceleration parameter $q$ increases, $N_{D(ABC)}$ decreases monotonically and eventually undergoes a sudden entanglement death, whereas $N_{C(ABD)}$ displays a significantly higher degree of robustness, vanishing only in the infinite-acceleration limit. This clear contrast indicates that the entanglement characterized by $N_{C(ABD)}$ is substantially more resilient to Unruh-induced thermal noise than that associated with $N_{D(ABC)}$.
Consequently, although both $N_{C(ABD)}$ and $N_{D(ABC)}$ can serve as useful entanglement resources in the low and moderate-acceleration regimes, their effectiveness is severely diminished at high accelerations. This asymmetric degradation of quantum entanglement highlights the crucial role of observer motion and reference-frame dependence in relativistic quantum information. It further emphasizes that a careful choice of subsystem partitioning is essential when designing quantum communication or quantum information-processing protocols involving non-inertial participants, in order to maximize the robustness and reliability of quantum correlations.

\begin{figure}
\begin{minipage}[t]{0.5\linewidth}
\centering
\includegraphics[width=3.0in,height=5.2cm]{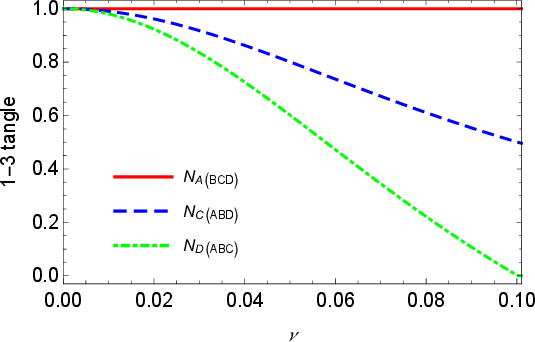}
\label{fig2}
\end{minipage}%
\caption{The $1-3$ tangle of the $CL_{4}$ state as a function of the effective coupling parameter $\nu$ with a fixed acceleration parameter $q=0.99$. }
\label{Fig2}
\end{figure}

In Fig.\ref{Fig2}, we examine the behavior of the $1-3$ tangle of the $CL_{4}$ state as a function of the effective coupling parameter $\nu$ in the extreme acceleration regime ($q=0.99$). This regime allows us to isolate the impact of detector-field interactions on quantum entanglement in non-inertial frames. Remarkably, the entanglement associated with inertial observers, quantified by $N_{A(BCD)}$ and $N_{B(ACD)}$, remains strictly maximal over the entire range of $\nu$, demonstrating complete immunity to variations in the coupling strength.
By contrast, the entanglement involving non-inertial subsystems exhibits qualitatively different behavior. The negativity $N_{C(ABD)}$, corresponding to the inertial observer Charlie, decreases monotonically with increasing $\nu$ but remains finite throughout the entire parameter range, never vanishing completely. In comparison, the negativity $N_{D(ABC)}$, associated with the accelerated observer David, also decreases monotonically as $\nu$ increases and undergoes a sudden entanglement death beyond a critical coupling strength.

Physically, this abrupt loss of entanglement---observed as relativistic effects intensify (whether due to the increasing acceleration $q$ in Fig.\ref{Fig1} or the effective coupling $\nu$ in Fig.\ref{Fig2})---can be understood as a direct consequence of the effective thermalization experienced by the accelerated detector $D$. Unlike the inertial observers, $D$ perceives the vacuum field as a thermal bath, where the effective temperature and interaction rate are governed by $q$ and $\nu$, respectively. This interaction acts as a local decoherence channel, introducing thermal noise that suppresses the off-diagonal elements of the density matrix responsible for entanglement. The partition $D(ABC)$ is uniquely sensitive to this effect because it isolates the single detector subjected to the Unruh noise from the remaining inertial system. Consequently, when the accumulated thermal mixedness exceeds a critical threshold, the negativity vanishes completely, resulting in the observed sudden death.

This pronounced disparity highlights the strong sensitivity of quantum entanglement to the participation of non-inertial observers. In particular, stronger detector-field coupling amplifies the degradation of quantum correlations involving accelerated subsystems. Our analysis indicates that this asymmetric entanglement evolution arises from the interplay between acceleration-induced Unruh thermal noise and the local field coupling strength: while acceleration establishes an effective thermal background, the coupling strength governs the rate and extent of entanglement decay. Overall, these results underscore the observer-dependent and asymmetric nature of entanglement dynamics in relativistic settings and further establish the $CL_{4}$ state as a promising and robust quantum resource in strongly non-inertial regimes.

\begin{figure}
\begin{minipage}[t]{0.5\linewidth}
\centering
\includegraphics[width=3.0in,height=6.24cm]{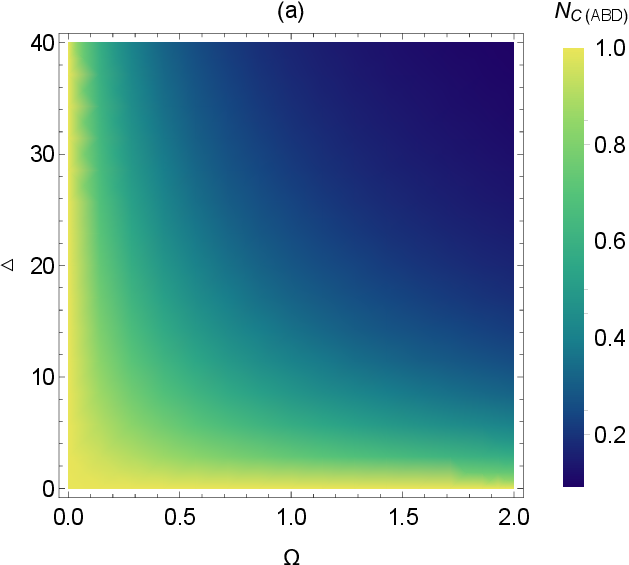}
\label{fig3a}
\end{minipage}%
\begin{minipage}[t]{0.5\linewidth}
\centering
\includegraphics[width=3.0in,height=6.24cm]{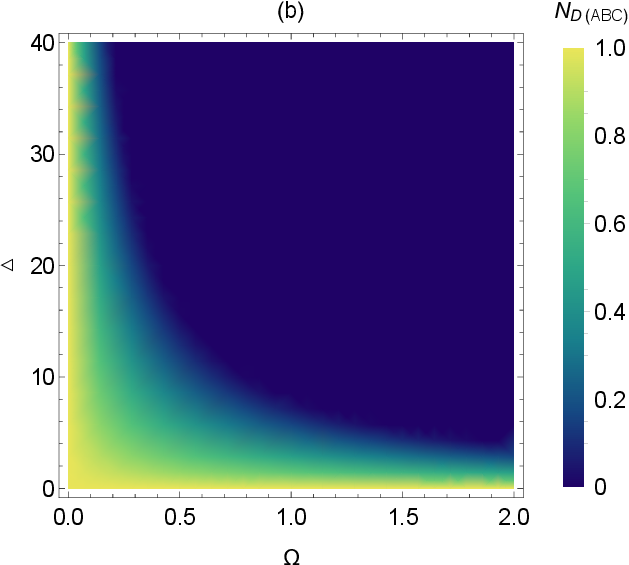}
\label{fig3b}
\end{minipage}%
\caption{The $N_{C(ABD)}$ and $N_{D(ABC)}$ of the $CL_4$ state as functions of the interaction time duration $\vartriangle$ and the energy gap $\Omega$, with $\epsilon^{2}=8\pi^{2}\cdot10^{-6}$, $\kappa=0.02$, and $q=0.9999$.  }
\label{Fig3}
\end{figure}

Fig.\ref{Fig3} illustrates the impact of the interaction between the accelerated detector and the external scalar field on the entanglement structure of the $CL_4$ state. Here, the negativities $N_{C(ABD)}$ and $N_{D(ABC)}$ are plotted as functions of the interaction duration $\Delta$ and the detector energy gap $\Omega$. Both negativities exhibit a monotonic decay with increasing $\Delta$ and $\Omega$; however, the degradation of $N_{D(ABC)}$ occurs substantially faster than that of $N_{C(ABD)}$, revealing a pronounced partition-dependent sensitivity to the Unruh effect.  This asymmetry highlights distinct operational regimes for inertial and
non-inertial subsystems and clarifies how interactions with field modes reshape the  quantum correlations. Moreover, the observed behavior indicates that the robustness of entanglement can be actively engineered by appropriately tuning the detector parameters. In particular, artificial two-level atoms with tailored energy gaps may mitigate Unruh-induced thermal noise, thereby helping to preserve nonclassical correlations. These findings provide a concrete strategy for maintaining reliable quantum information processing in relativistic or
non-inertial settings.

\section{Conclusions}
In this work, we have investigated the relativistic evolution of the tetrapartite $CL_4$ state using a physically realistic Unruh-DeWitt detector model. Contrary to the common expectation that the Unruh effect inevitably degrades initially maximal entanglement, we have shown that the entanglement between an inertial observer and the remaining tripartite subsystem remains strictly maximal across the entire acceleration range. Specifically, the entanglement retains its initial maximal value of unity, exhibiting a phenomenon we refer to as the \emph{complete freezing of initially maximal entanglement} under uniform acceleration. This behavior stands in sharp contrast to that of conventional quantum states such as Bell, GHZ, and W states, whose entanglement is progressively degraded or even destroyed by Unruh thermal noise \cite{SDF61,SDF62,SDF63,SDF64,SDF65}. The complete freezing observed here is not merely a numerical curiosity but carries significant implications for relativistic quantum information theory. It overturns the prevailing view that the Unruh effect universally diminishes initially maximal entanglement, demonstrating instead that certain highly structured $CL_4$ states possess intrinsic robustness against relativistic decoherence.  Our results thus provide the first explicit example of acceleration-invariant maximal entanglement within a fully relativistic detector framework, offering a novel perspective on the interplay between quantum correlations and non-inertial motion.

From a practical perspective, the \emph{complete freezing of initially maximal entanglement}  in the $CL_4$ state opens promising avenues for quantum technologies operating in relativistic regimes. Such states can serve as reliable resources for quantum communication and computation in curved spacetime or high-acceleration environments. The observed asymmetry in the entanglement of accelerated parties further underscores the importance of careful subsystem selection when designing relativistic quantum protocols. In summary, our work not only deepens the understanding of entanglement dynamics under the Unruh effect, but also identifies a new class of multipartite states capable of preserving initially maximal entanglement in extreme relativistic settings. These findings provide a foundation for further exploration of relativistically robust quantum resources and their potential applications in next-generation quantum technologies beyond inertial reference frames.

\begin{acknowledgments}
 This work is supported by the National Natural Science Foundation of China (12575056) and  the Special Fund for Basic Scientific Research of Provincial Universities in Liaoning under grant NO. LS2024Q002.	
\end{acknowledgments}


\end{document}